\documentstyle[twoside,epsf]{article}

\catcode`\@=11
\long\def\@makefntext#1{
\protect\noindent \hbox to 3.2pt {\hskip-.9pt  
$^{{\eightrm\@thefnmark}}$\hfil}#1\hfill}		

\def\thefootnote{\fnsymbol{footnote}}
\def\@makefnmark{\hbox to 0pt{$^{\@thefnmark}$\hss}}	
        
\def\ps@myheadings{\let\@mkboth\@gobbletwo
\def\@oddhead{\hbox{}
\rightmark\hfil\eightrm\thepage}   
\def\@oddfoot{}\def\@evenhead{\eightrm\thepage\hfil
\leftmark\hbox{}}\def\@evenfoot{}
\def\sectionmark##1{}\def\subsectionmark##1{}}

\oddsidemargin=\evensidemargin
\renewcommand{\thefootnote}{\fnsymbol{footnote}}

\newcounter{sectionc}\newcounter{subsectionc}\newcounter{subsubsectionc}
\renewcommand{\section}[1] {\vspace{12pt}\addtocounter{sectionc}{1} 
\setcounter{subsectionc}{0}\setcounter{subsubsectionc}{0}\noindent 
        {\tenbf\thesectionc. #1}\par\vspace{5pt}}
\renewcommand{\subsection}[1] {\vspace{12pt}\addtocounter{subsectionc}{1} 
        \setcounter{subsubsectionc}{0}\noindent 
        {\bf\thesectionc.\thesubsectionc. {\kern1pt \bfit #1}}\par\vspace{5pt}}
\renewcommand{\subsubsection}[1] {\vspace{12pt}\addtocounter{subsubsectionc}{1}
        \noindent{\tenrm\thesectionc.\thesubsectionc.\thesubsubsectionc.
        {\kern1pt \tenit #1}}\par\vspace{5pt}}
\newcommand{\nonumsection}[1] {\vspace{12pt}\noindent{\tenbf #1}
        \par\vspace{5pt}}

\newcounter{appendixc}
\newcounter{subappendixc}[appendixc]
\newcounter{subsubappendixc}[subappendixc]
\renewcommand{\thesubappendixc}{\Alph{appendixc}.\arabic{subappendixc}}
\renewcommand{\thesubsubappendixc}
	{\Alph{appendixc}.\arabic{subappendixc}.\arabic{subsubappendixc}}

\renewcommand{\appendix}[1] {\vspace{12pt}
        \refstepcounter{appendixc}
        \setcounter{figure}{0}
        \setcounter{table}{0}
        \setcounter{lemma}{0}
        \setcounter{theorem}{0}
        \setcounter{corollary}{0}
        \setcounter{definition}{0}
        \setcounter{equation}{0}
        \renewcommand{\thefigure}{\Alph{appendixc}.\arabic{figure}}
        \renewcommand{\thetable}{\Alph{appendixc}.\arabic{table}}
        \renewcommand{\theappendixc}{\Alph{appendixc}}
        \renewcommand{\thelemma}{\Alph{appendixc}.\arabic{lemma}}
        \renewcommand{\thetheorem}{\Alph{appendixc}.\arabic{theorem}}
        \renewcommand{\thedefinition}{\Alph{appendixc}.\arabic{definition}}
        \renewcommand{\thecorollary}{\Alph{appendixc}.\arabic{corollary}}
        \renewcommand{\theequation}{\Alph{appendixc}.\arabic{equation}}
        \noindent{\tenbf Appendix \theappendixc #1}\par\vspace{5pt}}
\newcommand{\subappendix}[1] {\vspace{12pt}
        \refstepcounter{subappendixc}
        \noindent{\bf Appendix \thesubappendixc. {\kern1pt \bfit #1}}
	\par\vspace{5pt}}
\newcommand{\subsubappendix}[1] {\vspace{12pt}
        \refstepcounter{subsubappendixc}
        \noindent{\rm Appendix \thesubsubappendixc. {\kern1pt \tenit #1}}
	\par\vspace{5pt}}

\newcommand{\textlineskip}{\baselineskip=13pt}
\newcommand{\smalllineskip}{\baselineskip=10pt}

\def\eightcirc{
\begin{picture}(0,0)
\put(4.4,1.8){\circle{6.5}}
\end{picture}}
\def\eightcopyright{\eightcirc\kern2.7pt\hbox{\eightrm c}} 

\newcommand{\copyrightheading}[1]
        {\vspace*{-2.5cm}\smalllineskip{\flushleft
        {\footnotesize International Journal of Modern Physics C, #1}\\
        {\footnotesize $\eightcopyright$\,\,\, World Scientific Publishing
         Company}\\
         }}

\def\abstracts#1#2#3{{
        \centering{\begin{minipage}{4.5in}\baselineskip=10pt\footnotesize
        \parindent=0pt #1\par
        \parindent=15pt #2\par
        \parindent=15pt #3\par
        \end{minipage}}\par}} 

\renewenvironment{thebibliography}[1]
        {\frenchspacing
	 \ninerm\baselineskip=11pt
         \begin{list}{\arabic{enumi}.}
        {\usecounter{enumi}\setlength{\parsep}{0pt}     
         \setlength{\leftmargin 17pt}{\rightmargin 0pt}   
         \setlength{\itemsep}{0pt} \settowidth
	{\labelwidth}{#1.}\sloppy}}{\end{list}}

\newcounter{itemlistc}
\newcounter{romanlistc}
\newcounter{alphlistc}
\newcounter{arabiclistc}

\newcommand{\fcaption}[1]{
        \refstepcounter{figure}
	\setbox\@tempboxa = \hbox{\footnotesize Fig.~\thefigure. #1}
	\ifdim \wd\@tempboxa > 5in
           {\begin{center}
	\parbox{5in}{\footnotesize\smalllineskip Fig.~\thefigure. #1}
            \end{center}}
        \else
             {\begin{center}
	     {\footnotesize Fig.~\thefigure. #1}
              \end{center}}
        \fi}

\newcommand{\tcaption}[1]{
        \refstepcounter{table}
	\setbox\@tempboxa = \hbox{\footnotesize Table~\thetable. #1}
        \ifdim \wd\@tempboxa > 5in
           {\begin{center}
         \parbox{5in}{\footnotesize\smalllineskip Table~\thetable. #1}
            \end{center}}
        \else
             {\begin{center}
	     {\footnotesize Table~\thetable. #1}
              \end{center}}
        \fi}

\def\@citex[#1]#2{\if@filesw\immediate\write\@auxout
	{\string\citation{#2}}\fi
\def\@citea{}\@cite{\@for\@citeb:=#2\do
	{\@citea\def\@citea{,}\@ifundefined
	{b@\@citeb}{{\bf ?}\@warning
	{Citation `\@citeb' on page \thepage \space undefined}}
	{\csname b@\@citeb\endcsname}}}{#1}}

\newif\if@cghi
\def\cite{\@cghitrue\@ifnextchar [{\@tempswatrue
	\@citex}{\@tempswafalse\@citex[]}}
\def\citelow{\@cghifalse\@ifnextchar [{\@tempswatrue
	\@citex}{\@tempswafalse\@citex[]}}
\def\@cite#1#2{{$\null^{#1}$\if@tempswa\typeout
	{IJCGA warning: optional citation argument 
	ignored: `#2'} \fi}}

\def\pmb#1{\setbox0=\hbox{#1}
        \kern-.025em\copy0\kern-\wd0
        \kern.05em\copy0\kern-\wd0
        \kern-.025em\raise.0433em\box0}

\def\fnt#1#2{\footnotetext{\kern-.3em
        {$^{\mbox{\scriptsize #1}}$}{#2}}}

\def\fpage#1{\begingroup
\voffset=.3in
\thispagestyle{empty}\begin{table}[b]\centerline{\footnotesize #1}
        \end{table}\endgroup}

\def\runninghead#1#2{\pagestyle{myheadings}
\markboth{{\protect\footnotesize\it{\quad #1}}\hfill}
{\hfill{\protect\footnotesize\it{#2\quad}}}}
\font\tenbf=cmbx10
\font\tenit=cmti10 
\font\tenit=cmti10
\font\bfit=cmbxti10 at 10pt

\font\ninerm=cmr9

\font\eightrm=cmr8

\def\lsym{\raise-3pt\hbox{\vbox{\tabskip0pt\offinterlineskip
	\halign{\tabskip0pt plus 1em
	##\tabskip0pt\cr
	$\,\,<\,\,$\cr
	$\,\,\sim\,\,$\cr}}}}
\def\rsym{\raise-3pt\hbox{\vbox{\tabskip0pt\offinterlineskip
     \halign{\tabskip0pt plus 1em
      ##\tabskip0pt\cr
      $\,\,>\,\,$\cr
      $\,\,\sim\,\,$\cr}}}}
\def\qed{\hbox{${\vcenter{\vbox{			
	\hrule height 0.4pt\hbox{\vrule width 0.4pt height 6pt
	\kern5pt\vrule width 0.4pt}\hrule height 0.4pt}}}$}}
\def\theequation{\thesection.\arabic{equation}}		
\renewcommand{\thefootnote}{\fnsymbol{footnote}}	

\def\l{\langle}
\def\r{\rangle}
\begin{document}

\runninghead{Y. Okabe, T. Miyajima, T. Ito \& T. Kawakatsu}
{Application of Monte Carlo Method to Phase Separation Dynamics \dots}

\normalsize\textlineskip
\thispagestyle{empty}
\setcounter{page}{1}

\copyrightheading{Vol. 0, No. 0 (1999) 000--000}

\vspace*{0.88truein}

\fpage{1}
\centerline{\bf APPLICATION OF MONTE CARLO METHOD TO PHASE}
\vspace*{0.035truein}
\centerline{\bf SEPARATION DYNAMICS OF COMPLEX SYSTEMS}
\vspace*{0.37truein}
\centerline{\footnotesize Yutaka OKABE$^*$, Tsukasa MIYAJIMA and Toshiro ITO} 
\vspace*{0.015truein}
\centerline{\footnotesize\it 
Department of Physics, Tokyo Metropolitan University,}
\baselineskip=10pt
\centerline{\footnotesize\it 
Hachioji, Tokyo 192-0397, Japan}
\vspace*{0.015truein}
\centerline{\footnotesize\it 
$^*$E-mail: okabe@phys.metro-u.ac.jp}
\vspace*{0.15truein}
\centerline{\footnotesize and}
\vspace*{0.15truein}
\centerline{\footnotesize Toshihiro KAWAKATSU} 
\vspace*{0.015truein}
\centerline{\footnotesize\it 
Department of Computational Science and Engineering,}
\vspace*{0.015truein}
\centerline{\footnotesize\it 
Nagoya University, Nagoya 464-8601, Japan}

\vspace*{0.225truein}

\vspace*{0.21truein}
\abstracts{
We report the application of the Monte Carlo simulation 
to phase separation dynamics.  First, we deal with 
the phase separation under shear flow.  
The thermal effect on the phase separation is discussed, 
and the anisotropic growth exponents 
in the late stage are estimated.
Next, we study the effect of surfactants on 
the three-component solvents.  We obtain the mixture of 
macrophase separation and microphase separation, 
and investigate the dynamics of both phase separations.
}{}{}



\vspace*{1pt}\textlineskip	
\section{Introduction}		
\vspace*{-0.5pt}
\noindent
The phase separation dynamics has attracted a lot of attention 
in recent decades.  Complex fluids, such as polymers, emulsions, 
and colloidal suspension, are of current interest.  
For simple systems it is considered that the domain-size growth 
in the late stage is governed by an algebraic law, $R(t) \sim t^n$. 
The classical Lifshitz-Slyozov theory\cite{Lifshitz61}
gives the growth exponent $n$=1/3 in the case of the spinodal 
decomposition of the conserved order parameter.  In contrast, 
the late-stage ordering process of the nonconserved order parameter 
is described by the classical Lifshitz-Allen-Cahn 
law\cite{Lifshitz62,Allen79}, $n=1/2$.  

However, the dynamics of more complex systems, such as 
the phase separation problem under shear flow, is not simple, 
and more attention has been given to these systems. 
Computer simulation is a powerful method, 
and as a mesoscopic approach, the time-dependent 
Ginzburg-Landau (TDGL) model is frequently used 
for the simulation of phase separation.

In the present paper, we use the Monte Carlo simulation based on 
the lattice model instead.  We apply the Monte Carlo method 
to two problems; the phase separation under shear flow\cite{Onuki97} 
and the phase separation of three-component solvents with surfactants.  
In the former problem, we mainly focus on the late-stage dynamics.
In the latter problem, we pay attention to the mixture of 
the macrophase separation and microphase separation.  
It is to be noted that we often encounter the problem of 
slow dynamics in simulational studies of ordering process 
of phase separation.  We briefly make a remark on the method 
to overcome the slow dynamics.

\setcounter{section}{2}
\setcounter{equation}{0}
\section{Phase Separation under Shear Flow}
\noindent
The anisotropic domain growth has been observed for the phase 
separation under shear flow both 
experimentally\cite{Takebe88,Hashimoto88} and theoretically\cite{Ohta90}.  
It is interesting to study the anisotropic domain growth law 
in the late stage\cite{Lauger95,Corberi98}.
Here, we develop a new simulation method to study the domain growth 
under shear flow based on the lattice model.  
We can systematically study the effect of thermal fluctuations 
on phase separation for a wide range of temperatures including
the critical temperature without any particular assumption.  

\setcounter{footnote}{0}
\renewcommand{\thefootnote}{\alph{footnote}}
%

%
\begin{figure}[tbh]
\epsfxsize=7.5cm 
\centerline{\epsfbox{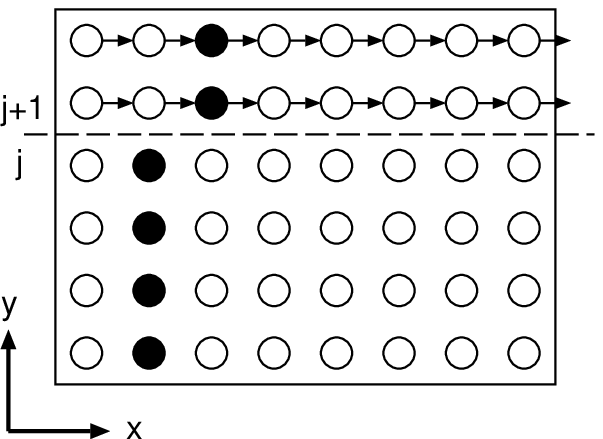}}
\vspace*{10pt}
\fcaption{The schematic illustration of the shear process 
in the lattice model.}
\end{figure}
%
We treat the two-dimensional model, and consider 
the case that the velocity field is given by 
\begin{equation}
v_x = \gamma y, \quad v_y = 0
\end{equation}
where $\gamma$ is the shear rate.  
In the lattice model, it corresponds to 
the relative slide of adjacent layers with one lattice spacing 
at the rate of $\gamma$ in unit time, which is schematically shown 
in Fig.~1.  Using this fact, we introduce shear flow in 
the Monte Carlo simulation of the phase separation dynamics. 
A similar idea was suggested by Chan\cite{Chan90}.
We employ the Kawasaki nearest-neighbor pair exchange 
for the Ising model because the system with the conserved order-parameter 
is assumed for the phase separation problem. 
The Hamiltonian of our system is given by 
\begin{equation}
{\cal H} = - J \sum_{\l i,j \r} \sigma_i \sigma_j 
\end{equation}
with the Ising coupling $J$.  
After the usual Metropolis spin-update process of 
one Monte Carlo step per spin, we perform the shear-flow process; 
we pick up a pair of adjacent layers to make a slide with 
the probability of $\gamma N_{{\rm layer}}$, where $N_{{\rm layer}}$ 
is the number of the layers.  It is to be noted that 
the Lees-Edwards boundary condition\cite{Lees72} is naturally demanded 
with the introduction of such a shear-flow process. 

We have made simulations for the square lattice, 
and the typical system size is $256 \times 256$ and $128 \times 128$. 
The volume fraction of each of the two components is chosen to be 0.5.  
We start from a random configuration, and quench a system to certain 
lower temperature.  We change the shear rate $\gamma$ and 
the quenching temperature to study the thermal effect 
on the phase separation.  

%
\begin{figure}[htb]
\epsfxsize=7.5cm 
\centerline{\epsfbox{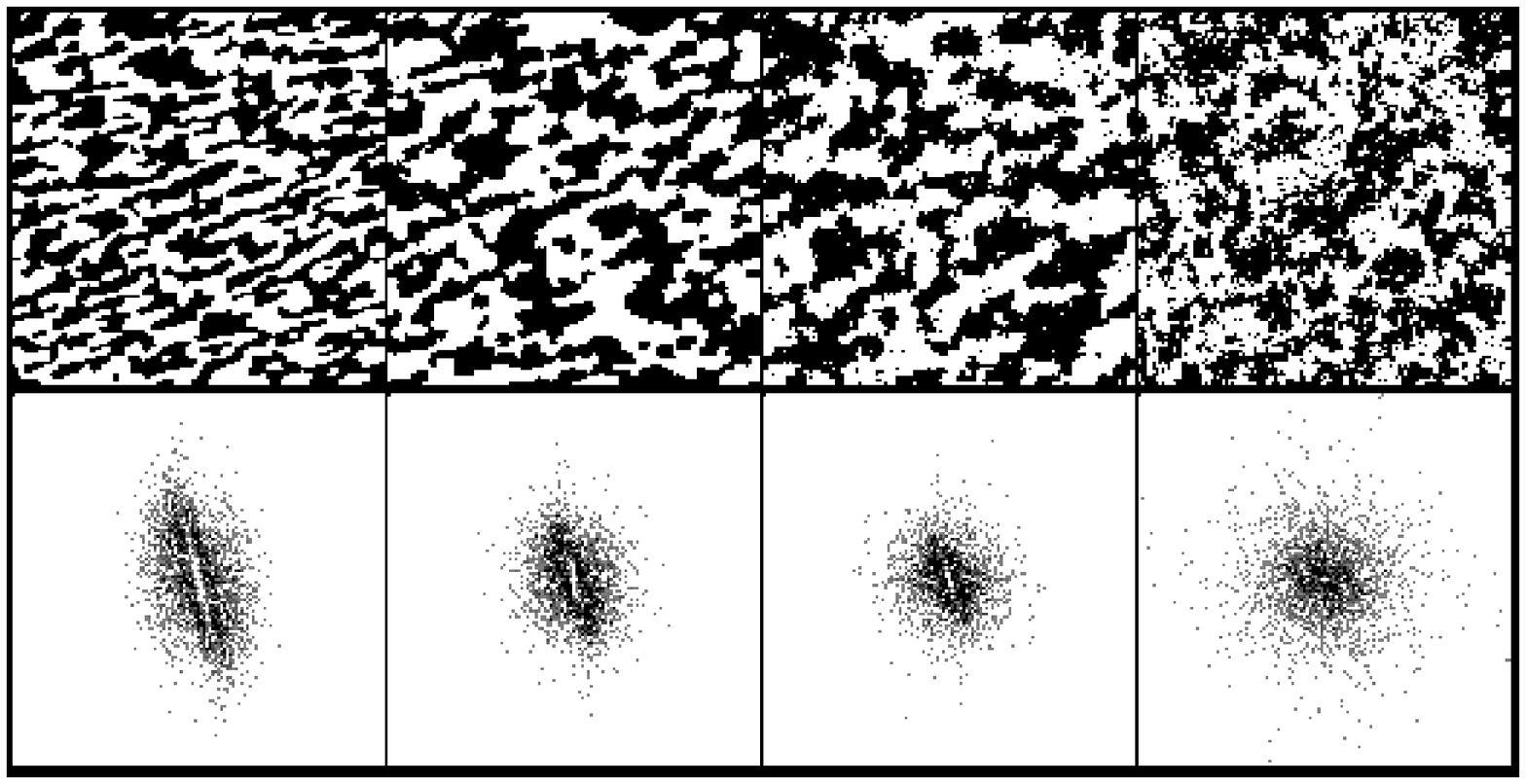}}
\vspace*{10pt}
\epsfxsize=7.5cm 
\centerline{\epsfbox{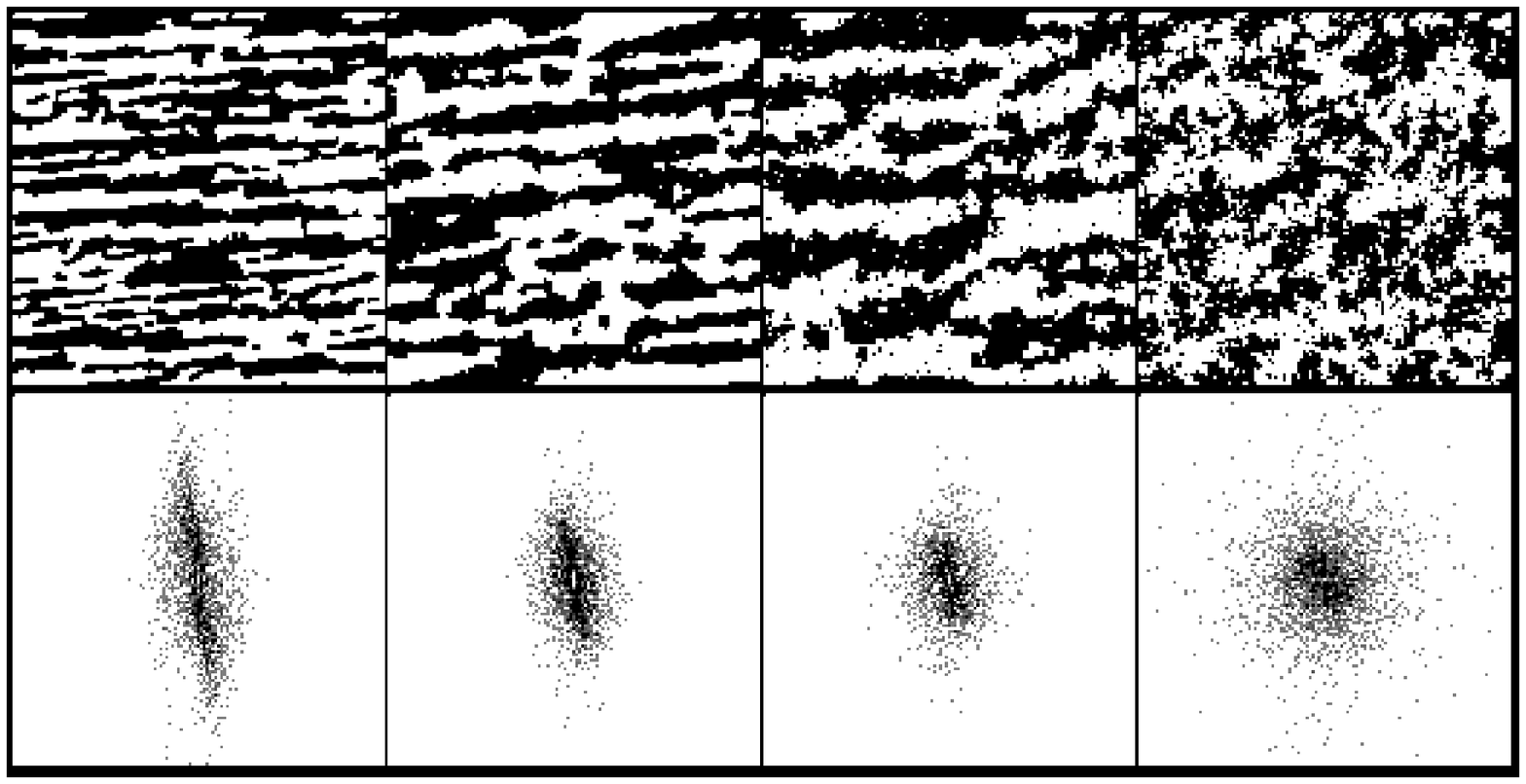}}
\vspace*{10pt}
\fcaption{The real-space snapshots (top) and corresponding structure 
factors (bottom).  The system size is $128 \times 128$.  
The shear rate $\gamma$ is 0.001, and the time is (a) 5000 MCS 
and (b) 20000 MCS.  The quenching temperatures are $T$=0.6, 1.2, 
1.8, and 2.4 from left to right.}
\end{figure}
%
Performing the fast Fourier transform, 
we calculate the structure factor $S({\vec k})$. 
We note that the Lees-Edwards boundary condition 
should be taken into account when performing the Fourier transform.  
This is the same situation 
as the study of the equilibrium properties of the Ising model 
with the tilted boundary condition\cite{Okabe98}. 
Examples of the real-space snapshots and corresponding structure 
factors are given in Fig.~2, where the system size is 
$128 \times 128$.  The shear rate $\gamma$ is 0.001, 
and the time is 5000 MCS and 20000 MCS for Fig.~2(a) and (b), 
respectively.  The quenching temperatures are 
$T$=0.6, 1.2, 1.8, and 2.4 from left to right.  We measure 
the temperature in unit of $J$.
We note that the transition temperature of the present system 
without shear flow is the Ising critical temperature, $T_c = 2.269 \cdots$.
From the figures, we see that below $T_c$ the shape of domain is elongated 
in the $x$ direction due to the effect of shear as time grows, 
which results in the break of the symmetry between $k_x$ and $k_y$.  
If there is no shear flow, the domain growth becomes very slow 
at low temperatures.  It is because the thermal diffusion is 
not so frequent there.  In contrast, the effect of shear becomes 
prominent at very low temperatures.  
There is a rapid elongation of the domain in the direction 
of shear flow.  At higher temperatures, even if the domain
elongates, the string-like domain breaks again because of 
thermal fluctuations.  Thus, thermal fluctuations play a role 
of hindering the effect of shear.

Let us consider the anisotropic growth in the late stage. 
If we assume that the structure factor $S({\vec k})$ has the elliptic 
symmetry as suggested from Fig.~2, we can calculate the slope of 
the principal axis, and the moments of $k$ 
along principal and subsidiary axes.  
We plot the time evolution of the second moments $k_+$ and $k_-$, 
which are the moments along the principal and subsidiary axes, 
respectively, in Fig.~3. We use the logarithmic scales, 
and the data for various shear rates, that is, 
$\gamma=0.00025, 0.001$ and $0.004$, are plotted in the same figure.  
We have chosen $\gamma t$ as the horizontal axis. 
The quenching temperature is 0.6, and the average has been 
taken over 16 samples for each $\gamma$. 
From the figure, we see how the domain deforms as time grows; 
$\l k_+^2 \r$ and $\l k_-^2 \r $ start to separate. 
%
\begin{figure}[htb]
\epsfxsize=8.5cm 
\centerline{\epsfbox{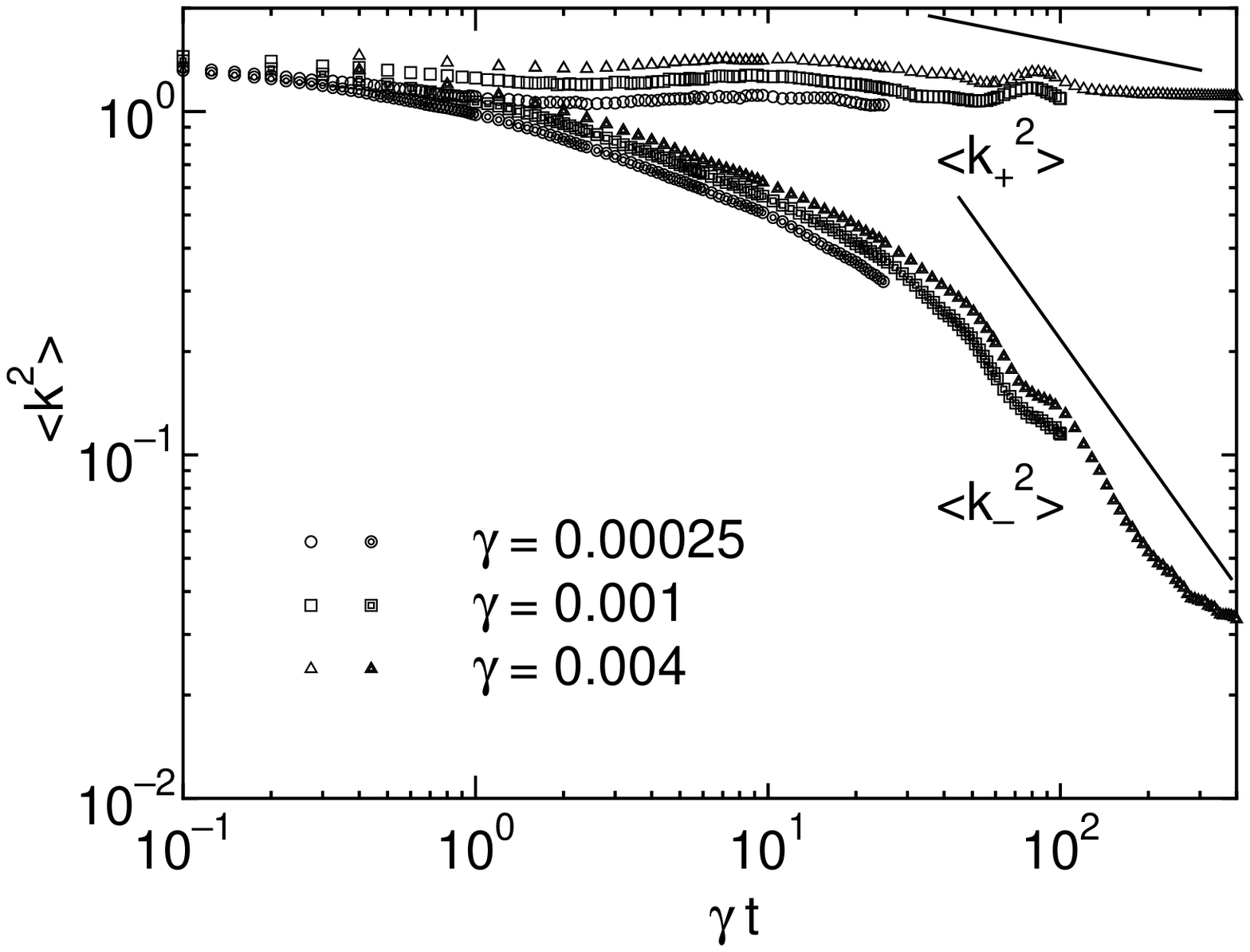}}
\vspace*{10pt}
\fcaption{The logarithmic plot of $\l k_{\pm}^2 \r$ as a function 
of $\gamma t$. The slopes of the straight lines are -1.4 and -0.2.}
\end{figure}

The anisotropic growth exponents in the late stage are recently 
discussed by Corberi, Gonnella, and Lamura\cite{Corberi98} 
using the TDGL model.  They have used the one-loop approximation 
for the TDGL model.  In our notation, the domain growth 
in the late stage is denoted by
\begin{eqnarray}
\l k_-^2 \r \sim t^{-2\alpha_x},  \hspace{1cm} t \rightarrow \infty
\\
\l k_+^2 \r \sim t^{-2\alpha_y},  \hspace{1cm} t \rightarrow \infty
\end{eqnarray}
with the anisotropic growth exponents $\alpha_x$ and $\alpha_y$.
In the late stage, the subsidiary axis $k_-$ for $S({\vec k})$ 
corresponds to the direction of elongation $x$ in real space.
The growth exponent is given by 1/3 for the system without shear 
flow\cite{Lifshitz61}.
We may estimate from the slope of the straight lines in Fig.~3 that
$\alpha_x=0.7$ and $\alpha_y=0.1$.  These values are smaller 
than the prediction of Ref.~9, that is, 
$\alpha_x=5/4$ and $\alpha_y=1/4$.  This is because the one-loop 
approximation employed in Ref.~9 neglects 
nonlinear mode-coupling effect, which is important in the 
late stage dynamics.  Quite recently, Corberi, Gonnela and Lamura 
have reported a new calculation using a renormalization group 
approach.  Their new values of $\alpha$'s are 
$\alpha_x=4/3$ and $\alpha_y=1/3$, which are still different from 
our numerical estimates. 
The details of the present calculation will be published 
elsewhere\cite{Miyajima99}.

\section{Three-Component Solvents with Surfactants}
\noindent
The effects of surfactants in binary mixtures is 
an interesting problem.  It produces various phases.  
The growth of domain size is hampered, 
and the microphase separation occurs as in the case of 
block copolymers.
The lattice model has been used to investigate the effects of 
surfactants\cite{Larson92,Stauffer94}.  The dynamics of 
the surfactant systems has been also studied 
using the lattice model by the Monte Carlo 
simulation\cite{Bernardes96,Larson96}.
%
\begin{figure}[htb]
\epsfxsize=5.5cm 
\centerline{\epsfbox{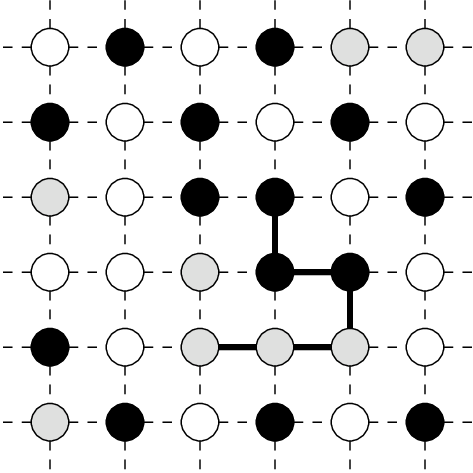}}
\vspace*{10pt}
\fcaption{Illustration of the lattice model of solvents with surfactants.}
\end{figure}
%

Here, we study the phase separation of three-component solvents 
with surfactants.  We can obtain a variety of interesting 
phases by increasing the number of components from two to three.
We consider the mixture of macrophase separation and microphase
separation for three-component systems. 
We use the Monte Carlo simulation to study the lattice model. 
Three-component solvents are represented by the three-state 
(A, B, C) Potts spins.  We represent the surfactants 
by diblock copolymer chains which are composed of two states, 
that is, A-B, B-C, or C-A.  We illustrate our model in Fig.~4. 
In this case, the segment number of two states in the diblock copolymer 
are 3 and 3.  We assume that the interaction between 
the solvent particles and that between the solvent particle 
and the monomer in the surfactant are the same. 
We use the Kawasaki pair exchange for solvents, and 
slithering-snake motions for the move of surfactants. 
These update processes are shown in Fig.~5.  
%
\begin{figure}[htb]
\epsfxsize=8.5cm 
\centerline{\epsfbox{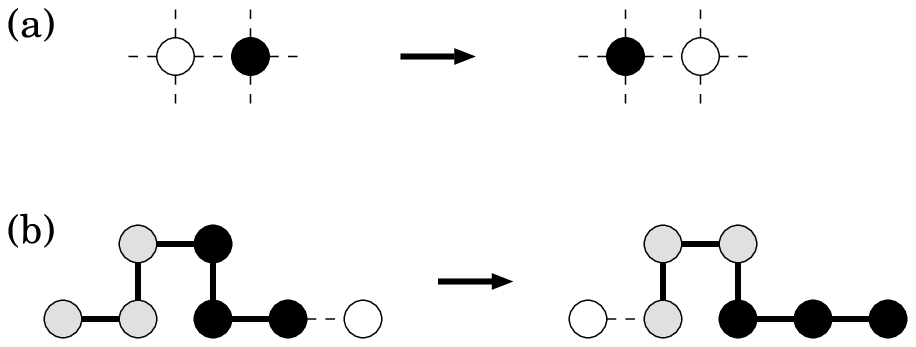}}
\vspace*{10pt}
\fcaption{Monte Carlo update processes; (a) Kawasaki pair exchange and 
(b) slithering-snake motion.}
\end{figure}

Our system suffers from the problem of slow dynamics, especially 
because of the existence of surfactants.  
To accelerate the phase separation, we use the trick of the 
replica-exchange Monte Carlo method\cite{Hukushima96a}.  
Although the dynamics including the replica-exchange process is 
not a simple one, it has been shown 
that the replica-exchange Monte Carlo method works well 
even for the estimate of the growth exponent
in the case of ordering process with an algebraic 
growth law\cite{Okabe99b}.

\begin{figure}[htb]
\epsfxsize=7.0cm 
\centerline{\epsfbox{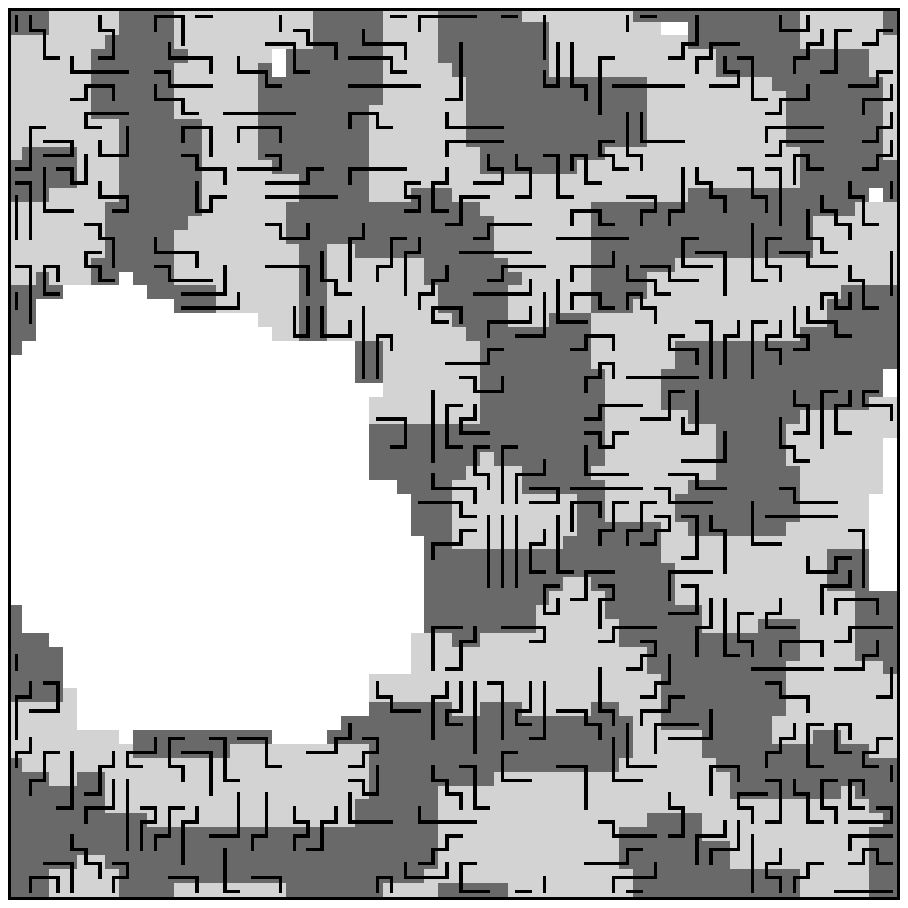}}
\vspace*{10pt}
\fcaption{Snapshot of the three-component system with surfactants.  
Only one type of surfactants, A-B, are included, 
which are represented by lines.
A and B particles are represented by black and gray 
squares, and C particles by white ones. }
\end{figure}
%
We have made simulations for several cases changing the volume fractions 
of solvents and surfactants.  Here, we show the result of 
one case.  We put only one type of surfactants, say A-B, 
in the three-component solvents.  The number of solvent particles 
for A, B and C are chosen as the same.  Starting from the random 
configuration, we quench the system at low temperature.
A typical snapshot in the late stage is given in Fig.~6.  
A and B particles are represented by black and gray 
squares, and C particles by white ones.  We observe the 
microphase separation of A and B particles.  We also see 
the macrophase separation of C particles and the mixture of 
other particles.  Thus, we find the mixture of macrophase 
separation and microphase separation.  
We should note that similar phenomena of complex phase separation 
have been studied in copolymer-homopolymer mixtures\cite{Ohta95}.

\begin{figure}[htb]
\epsfxsize=8.5cm 
\centerline{\epsfbox{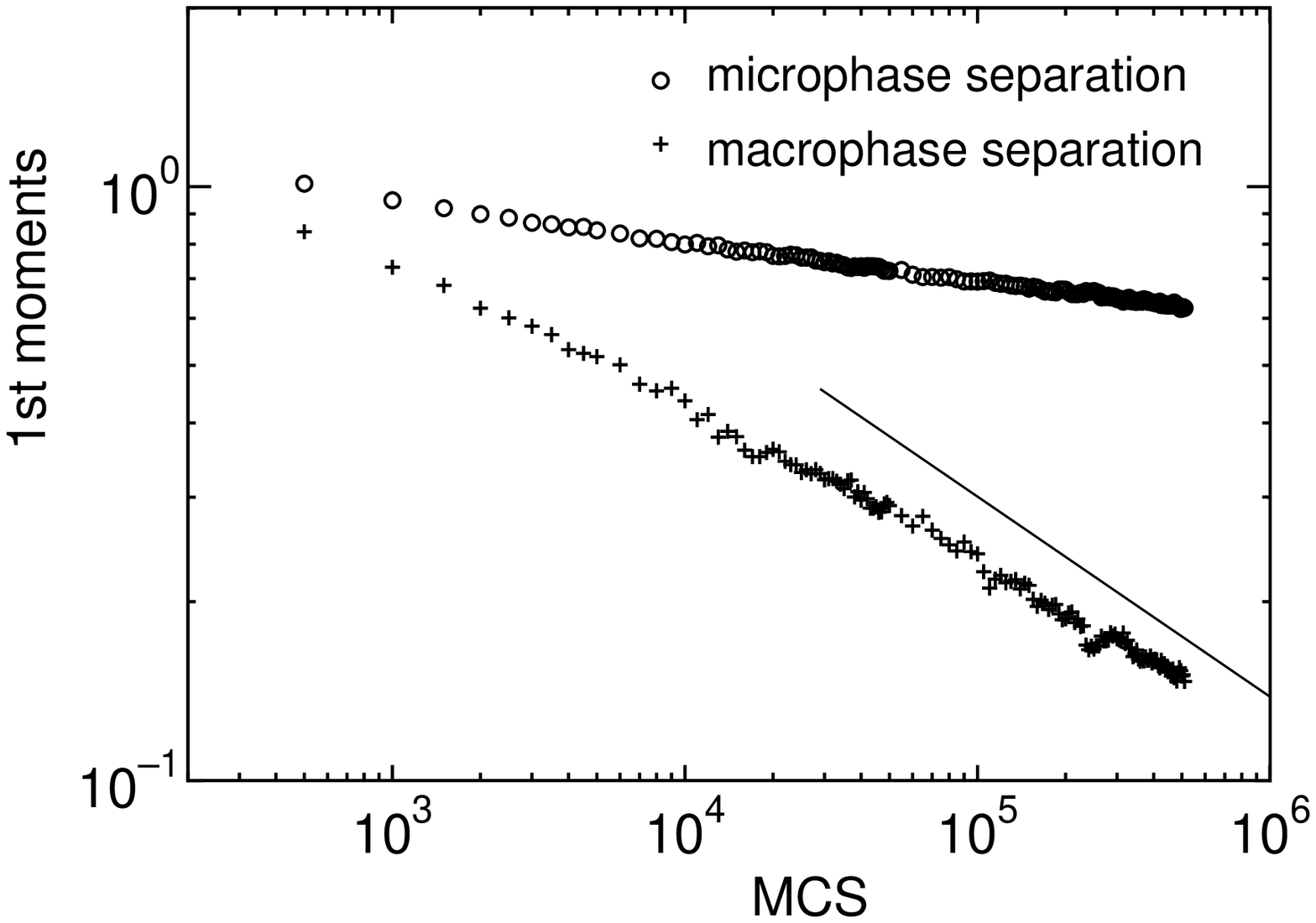}}
\vspace*{10pt}
\fcaption{Temporal growth of the first moment of the structure 
factors $S_({\vec k})$ which represent macrophase (CC) and 
microphase (AB) separations.  
The slope of the straight line is -1/3. }
\end{figure}
%
To study the dynamics of each phase separation process,
we calculate the structure factors, $S_{\rm CC}({\vec k}) = 
\l |\rho_{\rm C}({\vec k})|^2 \r$ and $S_{\rm AB}({\vec k}) = 
\l |\rho_{\rm A}({\vec k})-\rho_{\rm B}({\vec k})|^2 \r$, 
for macrophase separation 
and microphase separation, respectively.
Here, $\rho({\vec k})$ is the Fourier transform of density, and 
$\l \cdots \r$ denotes the thermal average.
The temporal evolution of the first moments 
of each structure factor are shown in Fig.~7.
We obtain the separate growth behavior for both phase separation 
processes.  The growth exponent of macrophase separation
is very close to the Lifshitz-Slyozov value\cite{Lifshitz61} 
of 1/3, whereas the microphase separation is much slower.
We will report the details of the present study 
elsewhere\cite{Ito99}.

\section{Summary}
\noindent
We have reported the Monte Carlo studies of two problems 
concerning the phase separation dynamics.  The first one is 
the phase separation dynamics under shear flow.  
Developing a new Monte Carlo method to study the phase 
separation dynamics under shear flow,  
we have studied the thermal effect on the phase separation. 
We have also discussed the anisotropic growth exponents 
in the late stage.
The second problem is the effect of surfactants on the 
three-component solvents.  We have obtained a mixture of 
macrophase separation and microphase separation, 
and have discussed the dynamics of these phase separations.

\nonumsection{Acknowledgments}
\noindent
This work was supported by a Grant-in-Aid for Scientific Research 
from the Ministry of Education, Science, Sports and Culture, Japan.
The computation in this work has been done using the facilities of 
the Supercomputer Center, Institute for Solid State Physics, 
University of Tokyo. 

\nonumsection{References}

\end{document}